\documentclass[useAMS,usenatbib]{mn2e}
\usepackage{graphicx}

\title[Optical IC emission from clusters]
{Optical inverse-Compton emission from clusters of galaxies}
\author[Yamazaki and Loeb]
{Ryo Yamazaki$^{1,2}$\thanks{E-mail: ryo@phys.aoyama.ac.jp}
and
Abraham Loeb$^{2}$
\\
$^{1}$Department of Physics and Mathematics, Aoyama Gakuin University, 
5-10-1, Fuchinobe, Sagamihara 252-5258, Japan \\
$^{2}$Harvard-Smithsonian Center for Astrophysics, 60 Garden Street, Cambridge, MA, 02138, USA
\\
}

\begin{document}


\pagerange{\pageref{firstpage}--\pageref{lastpage}} \pubyear{2015}
\maketitle
\label{firstpage}


\begin{abstract}

Shocks around clusters of galaxies accelerate electrons which
upscatter the Cosmic Microwave Background photons to higher-energies.
We use an analytical model to  calculate 
this inverse Compton (IC) emission, taking into account the effects of
additional energy losses via synchrotron and Coulomb scattering.
We find that the surface brightness of the optical IC emission 
increases with redshift and  halo mass. 
The IC emission surface brightness,
32--34~mag~arcsec$^{-2}$, for massive clusters is potentially
detectable by the newly developed Dragonfly Telephoto Array.

\end{abstract}



\section{Introduction}

According to the standard model of hierarchical structure formation,
accretion shocks occur around the virial radii of massive clusters
\citep[e.g.,][and references therein]{ryu2003,schaal2015}.
In these shocks, electrons are expected to be accelerated by 
first-order Fermi mechanism \citep[see, e.g.,][for recent review]{brunetti2014}
to have power-law distribution
\citep{blandford1978,bell1978}.
Turbulence in the intergalactic medium 
\citep[IGM;][]{ryu2008,takizawa2008,miniati2015} is another source of
particle acceleration via second-order Fermi mechanism
\citep{schlickeiser1987,brunetti2001,petrosian2001,fujita2003}.
Relativistic electrons 
give rise to  inverse Compton (IC) emission
by upscattering Cosmic Microwave Background (CMB) photons 
as well as radio synchrotron emission.
So far,  IC emission has been studied mainly in hard X-ray and gamma-ray bands 
both theoretically and observationally
\citep{sarazin1999,loeb2000,totani2000,takizawa2000,fujita2001,takizawa2002,
keshet2003,keshet2004a,keshet2012,petrosian2008,kushnir2010,bartels2015}.
However, currently there are only upper limits \citep[][and references therein]{ota2014,gastaldello2015}
and a few claimed detections of hard X-rays
\citep{rephaeli2008}, but
no detection in the gamma-ray band \citep{ackermann2010,ackermann2014},
which may constrain particle acceleration mechanisms
\citep[e.g.,][]{zandanel2014,vazza2015}.
Although IC emission from individual clusters has not yet been detected,
the cumulative emission from all of them
may contribute to extragalactic gamma-ray background 
\citep{loeb2000,miniati2002}.

In this paper, we focus on the IC emission namely in the optical band. 
So far, it has been thought that the optical IC emission is too dim  to be
detectable \citep[e.g.,][]{sarazin1999,fujita2001}. 
However, an advanced technique 
has recently been developed in the form of the
Dragonfly Telephoto Array \citep{abraham2014,vanDokkum2014},  
which is optimized for the detection
of extended ultra low surface brightness structures 
and is capable of imaging extended structures to surface brightness 
levels below 32~mag~arcsec$^{-2}$ in the SDSS g band with a reasonable exposure time.
This newly developed technique provides a new motivation for calculating 
the brightness of the optical IC emission in detail.
 Since no firm detection of IC emission at any wavelength has been reported as of yet,
 the optical telescopes hold the potential to bring the first clear detection of the IC
 emission from large-scale shocks around clusters of galaxies,
which is also the first evidence of nonthermal processes at accretion shocks.

We construct a simple one zone, analytical model for IC emission from cluster
shocks,
allowing us to capture the essential physical details as well as the parameter dependence
of the results.
Because the electrons emitting the optical IC emission have 
the Lorentz factor $\sim50$, they are potentially
affected by Coulomb energy losses \citep[e.g.,][]{sarazin1999,petrosian2008}.
However, as we show later,  this effect is not significant.
Our model applies also to other sources of nonthermal electrons.
We focus on IC from primarily accelerated electrons for simplicity,
although secondary electrons may also contribute 
\citep[e.g.,][]{blasi1999,miniati2003,inoue2005,kushnir2009}.
For the small magnetic field strength expected in clusters, the contribution of
synchrotron emission to the optical brightness is negligible.

Our paper is organized as follows. 
In section~2, we describe our analytical model for calculating the IC emission.
The surface brightness in the SDSS g-band is then calculated for fiducial parameters
in section~3. 
Finally, we summarize our results and predictions for specific clusters in section~4.
Throughout the paper, we assume a flat $\Lambda$CDM universe 
with cosmological parameters,
$h=0.6774$, $\Omega_{\rm m}=0.3089$,
$\Omega_\Lambda=0.6911$,  and 
$\Omega_{\rm b}=0.0486$
\citep{planck2015} 


\section{Analytical model of IC spectrum}
\label{sec:model}

\subsection{Physical quantities of cluster shocks}
A halo of mass $M$ collapsing at redshift $z$ has a virial radius $r_{\rm vir}$,
within which the mean density is $\Delta_{\rm c}$ times the critical density
$\rho_{\rm c}(z)$,
and circular velocity at the virial radius $V_{\rm c}$ given by \citep{bryan1998,barkana2001}, 
\begin{eqnarray}
r_{\rm vir}&=& 
\left( \frac{3M}{4\pi\Delta_{\rm c} \rho_{\rm c}(z)}\right)^{1/3}
\nonumber\\
&=&0.79~h^{-2/3}
M_{14}^{1/3}w(z)^{-1/3}(1+z)^{-1} {\rm Mpc}~~,
\label{eq:Rvir}
\end{eqnarray}
\begin{eqnarray}
V_{\rm c} &=& \sqrt{\frac{GM}{r_{\rm vir}}} \nonumber \\
&=& 7.4\times10^7 h^{1/3}M_{14}^{1/3}
w(z)^{1/6}(1+z)^{1/2} {\rm cm}~{\rm s}^{-1}~~,
\label{eq:circ}
\end{eqnarray}
where $M_{14}=(M/10^{14}M_{\sun})$.  
The function $w(z)$ is given by
\begin{equation}
w(z)=\frac{\Omega_{\rm m}}{\Omega_{\rm m}^z}
\frac{\Delta_{\rm c}}{18\pi^2} ~~,
\end{equation}
where
\begin{equation}
\Omega_{\rm m}^z = \frac{\Omega_{\rm m}(1+z)^3}
{\Omega_{\rm m}(1+z)^3+\Omega_\Lambda}~~,
\end{equation}
\begin{equation}
\Delta_{\rm c}=18\pi^2+82(\Omega_{\rm m}^z-1)-39(\Omega_{\rm m}^z-1)^2~~.
\end{equation}
%
Note that $w(z)$ is a monotonically decreasing function of $z$,
starting at $w(0)=0.58$, decaying through $w(1)=0.35$ 
and asymptotically approaching 
$w(z)\rightarrow\Omega_{\rm m}=0.31$ as $z\rightarrow\infty$.

For simplicity, we assume that a spherical virial shock is formed at $r_{\rm vir}$,
and that accretion is smooth and not associated with mergers of sub-units.
Recent numerical simulations have shown that accretion shocks are deformed and
far from spherical \citep[e.g.,][]{ryu2003,lau2014,schaal2015,nelson2015}, 
however, emsemble-averaged gas profile shows that
virial shocks exist near the virial radius.
Future extensions of this work can be based on numerical simulations of non-spherical  configurations. 
The ensemble-averaged mass accretion rate onto the halo
is written as \citep{white1994},
\begin{equation}
\dot{M}=f_{\rm acc}\frac{V_{\rm c}^3}{G}~~,
\label{eq:Mdot}
\end{equation}
and the shock temperature is given by,
\begin{equation}
kT=f_{\rm T}\mu m_p V_{\rm c}^2 ~~,
\label{eq:temp}
\end{equation}
where $\mu m_p$ is the average mass of a particle (including electrons),
and we adopt $\mu=0.6$.
The factors $f_{\rm acc}$ and $f_{\rm T}$ are dimensionless
numbers of order unity.
%
The gas density just in front of the shock can be written as,
\begin{eqnarray}
n_g &=&\frac{(\Omega_{\rm b}/\Omega_{\rm m})\dot{M}}
  {4\pi r_{\rm vir}^2V_{\rm c}\mu m_p} \nonumber \\
  &=& 8.0\times10^{-5}f_{\rm acc}w(z) (1+z)^3~{\rm cm}^{-3}~~.
\label{eq:density}
\end{eqnarray}
This estimate is roughly consistent with the results given by
\citet{patej2015} if $f_{\rm acc}\approx0.5$.
The gas is compressed at the shock with shock compression ratio $r$, which
is somewhat uncertain.
Recent numerical simulations have shown that the shocks are not so strong
and their typical Mach number is around a few \citep{ryu2003,schaal2015},
i.e., $r<4$.
The IGM is pre-heated and the shocks are not so strong
at present epoch $z\sim0$, however, at high redshifts $z\ga1$ when the IGM is cold, 
the shocks around clusters are likely to be strong ($r\approx4$).
Fortunately, we will see in section~3 that the IC flux does not depend on $r$
for fiducial parameters.

The magnetic field around the shock is amplified as in supernova remnants
\citep[e.g.,][]{vink2003,bamba2003,bamba2005a,bamba2005b}.
Assuming that the energy density of the downstream magnetic field
constitutes a fraction $\xi_{\rm B}$ of the downstream thermal energy density,
we estimate the magnetic field strength as 
\citep{waxman2000,keshet2004b,fujita2005,kushnir2009},
\begin{eqnarray}
B&=&(12\pi r\xi_{\rm B}n_gf_{\rm T}\mu m_p )^{1/2}V_{\rm c} \nonumber\\
&=&0.72 ~(r_4f_{\rm T}f_{\rm acc}\xi_{{\rm B},-2})^{1/2}
M_{14}^{1/3}w(z)^{2/3}(1+z)^2~\mu{\rm G}~~,
\label{eq:mag}
\end {eqnarray}
where $\xi_{{\rm B},-2}=(\xi_{\rm B}/0.01)$ and $r_4=(r/4)$.

\subsection{Injected electron spectrum}
We assume a single power-law form of injected  electrons
\begin{equation}
\dot{N}(\gamma) =N_0 \gamma^{-p}~~,~~ (\gamma_{\rm min}<\gamma<\gamma_{\rm max})~~,
\label{eq:injspec}
\end{equation}
with a constant normalization $N_0$ and a spectral index $p$.

The maximum Lorentz factor, $\gamma_{\rm max}$, is determined by the balance of acceleration time
and the cooling time.
The acceleration time is given by \citep{drury83},
\begin{equation}
t_{\rm acc}(\gamma) =
\frac{r(r+1)}{r-1}
\frac{\eta_g \gamma m_ec^3}{eBv_{\rm sh}^2}~~,
\label{eq:Tacc}
\end{equation}
where the gyro-factor $\eta_g$ is of order unity, and the shock velocity that is measured
in the rest frame of the shock, $v_{\rm sh}$, is
related to $V_{\rm c}$ through $v_{\rm sh}=[r/(r-1)]V_{\rm c}$.
Here we assume Bohm diffusion with no change in the diffusion properties across the shock,
and no shock modification due to accelerated particles.
We equate $t_{\rm acc}$ to the cooling time via synchrotron and IC emission,
\begin{equation}
t_{\rm IC/syn}(\gamma) = \frac{6\pi m_e c}{\sigma_{\rm T}(B^2+B_{\rm CMB}^2)\gamma}~~,
\label{eq:Tic}
\end{equation}
where $B_{\rm CMB}=3.24~(1+z)^2~\mu$G, to obtain \citep{loeb2000},
\begin{eqnarray}
\gamma_{\rm max} &=&3.4\times10^7
r_4^{3/4}\left(\frac{15}{r^2-1}\right)^{1/2} \nonumber\\
&& \ \ 
\times
\frac{(f_{\rm T}f_{\rm acc}\xi_{{\rm B},-2})^{1/4}M_{14}^{1/2}}
{\eta_g^{1/2}[1+(B/B_{\rm CMB})^2]^{1/2}}
\frac{w(z)^{1/2}}{(1+z)^{1/2}}~~,
\end{eqnarray}
based on Eqs.~(\ref{eq:circ}) and (\ref{eq:mag}).
We note that the effect of synchrotron cooling is negligible since,
\begin{equation}
\left(\frac{B}{B_{\rm CMB}}\right)^2 = 4.9\times10^{-2}
r_4 f_{\rm T}f_{\rm acc}\xi_{{\rm B},-2}M_{14}^{2/3}w(z)^{4/3}~~,
\label{eq:Bratio}
\end{equation}
is always small for our adopted parameters.

In order to determine both the normalization constant $N_0$ and 
the minimum Lorentz factor $\gamma_{\rm min}$,
we make two assumptions. 
One is that the production rate of the accelerated electrons 
is a fraction $\eta_e$ of the particle number input rate across the virial shock,
$\dot{N}_{\rm in}=(\Omega_{\rm b}/\Omega_{\rm m})\dot{M}/\mu m_p$.
The other is that a fraction $\xi_e$ of thermal shock energy
$(3/2)kT\dot{N}_{\rm in}$ is carried by relativistic electrons.
These conditions can be written as
\begin{equation}
\int_{\gamma_{\rm min}}^{\gamma_{\rm max}}\dot{N}(\gamma) d\gamma 
= \eta_e \dot{N}_{\rm in}~~,
\label{eq:fracnum}
\end{equation}
\begin{equation}
\int_{\gamma_{\rm min}}^{\gamma_{\rm max}}
\gamma m_ec^2\dot{N}(\gamma) d\gamma 
= \xi_e \frac{3}{2}kT\dot{N}_{\rm in}~~.
\label{eq:fracene}
\end{equation}
One can solve these two equations numerically 
for $N_0$ and $\gamma_{\rm min}$,
given $p$, $\gamma_{\rm max}$, $\eta_e$  and $\xi_e$.
For our fiducial parameter set, 
$\gamma_{\rm min}$ is much smaller than $\gamma_{\rm max}$
and if $p>2$ it is approximately given  by,
\begin{eqnarray}
\gamma_{\rm min}&\approx&\frac{3(p-2)kT}{2(p-1)\eta_em_ec^2}\nonumber\\
&=&8.4~\frac{(p-2)\xi_{e,-2}}{(p-1)\eta_{e,-5}}
f_{\rm T}M_{14}^{2/3}w(z)^{1/3}(1+z)~~,
\label{eq:gminan}
\end{eqnarray}
where $\xi_{e,-2}=(\xi_e/0.01)$ and
$\eta_{e,-5}=(\eta_e/10^{-5})$.
For our fiducial parameters, this approximate
formula is accurate enough as long as $p>2.2$.
Finally, for convenience, Eq.~(\ref{eq:fracene}) can be written as,
\begin{equation}
N_0m_ec^2 =\frac{\xi_e}{f(p)}\frac{3}{2}kT\frac{(\Omega_{\rm b}/\Omega_{\rm m})\dot{M}}
{\mu m_p}~~,
\label{eq:norm1}
\end{equation}
where
\begin{equation}
f(p)=
\left\{
\begin{array}{lc}
\ln(\gamma_{\rm max}/\gamma_{\rm min}) & (p=2) \\
\frac{1}{p-2}\gamma_{\rm min}^{2-p}[1-(\gamma_{\rm min}/\gamma_{\rm max})^{p-2}] & (p\ne2) \\
\end{array}
\right. ~~.
\end{equation}

\subsection{Spectrum of IC emission}

Next we derive analytically the radiation spectrum of the IC emission
for a power-law distribution of injected electrons given by Eq.~(\ref{eq:injspec}).
A similar analysis has been done for synchrotron radiation  in the study of gamma-ray bursts
\citep[e.g.,][]{sari1998}.

The radiation power and the characteristic frequency of upscattered CMB photons 
which is radiated from a relativistic 
electron with Lorentz factor $\gamma$ are \citep{blumenthal1970},
\begin{equation}
P(\gamma)=\frac{4}{3}\sigma_{\rm T}c \gamma^2 \frac{B_{\rm CMB}^2}{8\pi}~~,
\end{equation}
\begin{eqnarray}
\nu(\gamma)&=&\frac{4}{3}\gamma^2\bar{\nu}_{\rm CMB}\nonumber\\
&=& 2.05\times10^{11}\gamma^2(1+z)~{\rm Hz}~~,
\label{eq:chfreq}
\end{eqnarray}
where
$h\bar{\nu}_{\rm CMB}\approx2.70~kT_{\rm CMB}$ is the mean energy of CMB photons
and we adopt $T_{\rm CMB}=2.726~(1+z)$~K \citep{fixsen2009}.
The spectral power, $P_\nu$ (power per unit frequency, in units of 
ergs~s$^{-1}$Hz$^{-1}$), 
is proportional to $\nu$ for $\nu<\nu(\gamma)$ 
and cuts off sharply at $\nu>\nu(\gamma)$.
The function $P_\nu$ is peaked around $\nu(\gamma)$,
and its peak value is well approximated as
$P_{\nu,{\rm max}}\approx P(\gamma)/\nu(\gamma)$.
Note that $P_{\nu,{\rm max}}$ is independent of $\gamma$.

The above description of $P_\nu$ is only suitable when the electron 
does not lose a significant fraction of its energy to radiation.
This requires the characteristic cooling time of the electrons to be longer than
the dynamical time of a cluster, which is given by,
\begin{equation}
t_{\rm dyn}=\frac{r_{\rm vir}}{V_{\rm c}} 
 = 1.5~w(z)^{-1/2}(1+z)^{-3/2}~{\rm Gyr}~~.
 \label{eq:Tdyn}
 \end{equation}
 Otherwise, the effect of energy loss must be considered.
The energy loss rate of electrons is dominated by Coulomb collisions at low energies
and synchrotron and IC losses at high energies
\citep{sarazin1999,petrosian2008}.
The cooling time via Coulomb collisions is well approximated for relativistic electrons as,
\begin{equation}
t_{\rm Clmb}(\gamma)=\frac{2}{3\sigma_{\rm T}cn_e\ln\Lambda}\gamma~~.
\label{eq:Tclmb}
\end{equation}
In the following, we assume that the electron density downstream of the shock 
is $n_e\approx0.5rn_g$,
where the factor 0.5 represents the number fraction of electrons to total gas particles,
and the gas density $n_g$ is given 
in Eq.~(\ref{eq:density}).
For simplicity, we fix a Coulomb logarithm at a value  $\ln\Lambda=40$.
The synchrotron and IC cooling times have been already derived in
Eq.~(\ref{eq:Tic}).
One can find the Lorentz factors, $\gamma_{\rm b1}$, 
$\gamma_{\rm b2}$ and $\gamma_{\rm b3}$,
at which two of the three timescales,
$t_{\rm IC/syn}(\gamma)$, $t_{\rm IC/syn}(\gamma)$ and $t_{\rm dyn}$,
 are balanced, such that,
\begin{eqnarray}
t_{\rm IC/syn}(\gamma_{\rm b1}) &=& t_{\rm dyn}\\
t_{\rm Clmb}(\gamma_{\rm b2}) &=& t_{\rm IC/syn}(\gamma_{\rm b2})\\
t_{\rm Clmb}(\gamma_{\rm b3})&=& t_{\rm dyn}
\end{eqnarray}
Since $t_{\rm IC/syn}\propto\gamma^{-1}$,
$t_{\rm Clmb}\propto\gamma$ and $t_{\rm dyn}\propto\gamma^0$,
$\gamma_{\rm b2}$ is always between $\gamma_{\rm b1}$ and $\gamma_{\rm b3}$.
Electrons with Lorentz factor $\gamma$ do not suffer significant cooling only if 
 $\gamma_{\rm b3}<\gamma<\gamma_{\rm b1}$.

\begin{figure*}
\includegraphics[width=180mm]{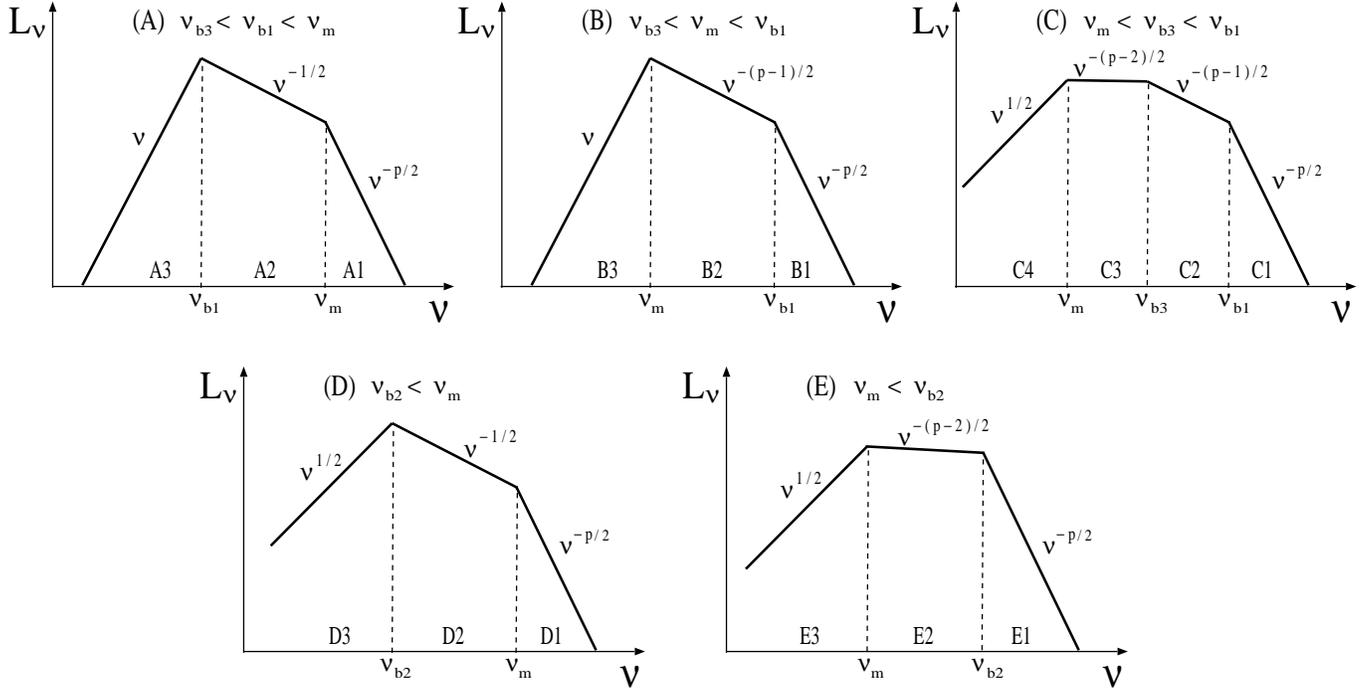}
\caption{
IC Spectrum from power-law distribution of relativistic electrons.
Regimes 
(A) $\nu_{\rm b3}<\nu_{\rm b1}<\nu_{\rm m}$, 
(B) $\nu_{\rm b3}<\nu_{\rm m}<\nu_{\rm b1}$
and (C) $\nu_{\rm m}<\nu_{\rm b3}<\nu_{\rm b1}$
are for $\nu_{\rm b3}<\nu_{\rm b2}<\nu_{\rm b1}$, 
while regimes 
(D) $\nu_{\rm b2}<\nu_{\rm m}$
and (E) $\nu_{\rm m}<\nu_{\rm b2}$
are for $\nu_{\rm b1}<\nu_{\rm b2}<\nu_{\rm b3}$.
}
\label{fig:spect}
\end{figure*}

To find the spectral shape of the IC emission taking into account
the electron cooling,
we define characteristic frequencies as,
\begin{equation}
\nu_{{\rm b}i} = \nu(\gamma_{{\rm b}i})~~,~~i=1,2,3
\end{equation}
and obtain,
\begin{eqnarray}
\nu_{\rm b1}&=&4.8\times10^{17}
\frac{w(z)(1+z)^{-4}}
{[1+(B/B_{\rm CMB})^2]^2}~{\rm Hz}~~,\\
\nu_{\rm b2}&=&2.9\times10^{15}
\left(\frac{\ln\Lambda}{40}\right)
\frac{r_4f_{\rm acc}w(z)}{1+(B/B_{\rm CMB})^2}
~{\rm Hz}~~,\\
\nu_{\rm b3}&=&1.8\times10^{13}
\left(\frac{\ln\Lambda}{40}\right)^2
r_4^2 f_{\rm acc}^2
w(z)(1+z)^4~{\rm Hz}~~.
\end{eqnarray}
These three frequencies coincide at a redshift $z_{\rm eq}$ approximately given by,
\begin{equation}
1+z_{\rm eq}\approx3.6
\left(\frac{\ln\Lambda}{40}\right)^{-1/4}
(r_4 f_{\rm acc})^{-1/4}~~,
\end{equation}
where the term $(B/B_{\rm CMB})^2$ is small and hence neglected
[see Eq.~(\ref{eq:Bratio})].
One can see $\nu_{\rm b3}<\nu_{\rm b2}<\nu_{\rm b1}$ for
$z<z_{\rm eq}$ and
$\nu_{\rm b1}<\nu_{\rm b2}<\nu_{\rm b3}$
for $z>z_{\rm eq}$.
The spectral shapes are different for these two cases, and are treated separately
in the following.
Our results are summarized in Figure~\ref{fig:spect}. 

\subsubsection{IC spectrum for $\nu_{\rm b3}<\nu_{\rm b2}<\nu_{\rm b1}$}

An electron with an initial Lorentz factor $\gamma>\gamma_{\rm b1}$
cools down to $\gamma_{\rm b1}$ in the dynamical time $t_{\rm dyn}$. 
Recall that the peak value of the instantaneous emissivity, $P_{\nu,{\rm max}}$,
is independent of the electron energy. Thus,
the average emission power at frequency $\nu$
is proportional to the cooling time of electrons with Lorentz factor $\gamma$,
satisfying $\nu=\nu(\gamma)\propto\gamma^2$. 
Therefore, the average spectrum scales as\footnote{
This spectral slope is identical to the similar case considered in
\citet{sari1998} since both of the characteristic frequencies of the synchrotron
and IC emissions are proportional to the square of the electron energy.}
 $P_{\nu}\propto t_{\rm IC/syn}(\gamma)\propto\nu^{-1/2}$
 for $\nu_{\rm b1}<\nu<\nu(\gamma)$.
For $\nu<\nu_{\rm b1}$, the spectrum has a low-energy tail, $P_\nu\propto\nu$.
For $\nu>\nu(\gamma)$, the spectrum steeply decays.
The averaged spectrum from such electrons has a peak at $\nu_{\rm b1}$.

An electron with an initial Lorentz factor $\gamma_{\rm b3}<\gamma<\gamma_{\rm b1}$
does not suffer significant cooling, and  the radiation spectrum is
$P_{\nu}\propto\nu$ for $\nu<\nu(\gamma)$ with a sharp cut off for
$\nu>\nu(\gamma)$.
An electron with an initial Lorentz factor $\gamma<\gamma_{\rm b3}$
suffer energy loss via Coulomb loss, so that we obtain
$P_\nu\propto t_{\rm Clmb}(\gamma)\propto\nu^{1/2}$
for $\nu<\nu(\gamma)$ with sharp cutoff for $\nu>\nu(\gamma)$.


To calculate the net spectrum from a power-law distribution
of electrons, one needs to integrate over $\gamma$. 
There are three
different cases, depending on $\gamma_{\rm min}$,
in which
(A)~$\gamma_{\rm b3}<\gamma_{\rm b1}<\gamma_{\rm min}$,
(B)~$\gamma_{\rm b3}<\gamma_{\rm min}<\gamma_{\rm b1}$ 
and
(C)~$\gamma_{\rm min}<\gamma_{\rm b3}<\gamma_{\rm b1}$.
Below we provide additional details on these regimes.

\begin{itemize}

\item
\underline{
(A)~$\gamma_{\rm b3}<\gamma_{\rm b1}<\gamma_{\rm min}$ 
(i.e., $\nu_{\rm b3}<\nu_{\rm b1}<\nu_{\rm m}$):}\\

In this  case, all the electrons cool down to $\gamma_{\rm b1}$.
The spectral power is given by,
\begin{equation}
\nu L_\nu =
\left\{
\begin{array}{ll}
L_0(\nu/\nu_{\rm m})^{(2-p)/2} & ({\rm A1}:~\nu_{\rm m}<\nu) \\
L_0(\nu/\nu_{\rm m})^{1/2} & ({\rm A2}:~\nu_{\rm b1}<\nu <\nu_{\rm m})\\
L_0(\nu_{\rm m}/\nu_{\rm b1})^{3/2}(\nu/\nu_{\rm m})^2 & ({\rm A3}:~\nu <\nu_{\rm b1})\\
\end{array}
\right. ~~,
\label{eq:LumiA}
\end{equation}
where,
\begin{equation}
\nu_{\rm m}=\nu(\gamma_{\rm min})~~.
\label{eq:nm}
\end{equation}
To determine the normalization constant, $L_0$, one can use the fact that
electrons with $\gamma>\gamma_{\rm b1}$ lose almost all their energy via
IC emission, that is, the luminosity in the regime  $\nu_{\rm m}<\nu$
can be written as \citep{loeb2000,keshet2003,kushnir2009},
\begin{equation}
L_\nu d\nu = \gamma m_ec^2\dot{N}(\gamma)d\gamma~~.
\end{equation}
Using Eqs.~(\ref{eq:injspec}) and (\ref{eq:norm1}) and 
$d\nu/\nu=2d\gamma/\gamma$ for $\nu\propto\gamma^2$,
we obtain,
\begin{eqnarray}
L_0&=&\frac{3\xi_ekT}{4g(p)}
\frac{(\Omega_{\rm b}/\Omega_{\rm m})\dot{M}}{\mu m_p}\nonumber \\
&=& 2.0\times10^{43}
\frac{f_{\rm T}f_{\rm acc}\xi_{e,-2}}{g(p)}M_{14}^{5/3}
w(z)^{5/6}(1+z)^{5/2}{\rm erg}~{\rm s}^{-1} ~~,
\label{eq:norm2}
\end{eqnarray}
where,
\begin{equation}
g(p)=
\left\{
\begin{array}{lc}
\ln(\gamma_{\rm max}/\gamma_{\rm min}) & (p=2) \\
\frac{1}{p-2}[1-(\gamma_{\rm min}/\gamma_{\rm max})^{p-2}] & (p\ne2) \\
\end{array}
\right. ~~.
\end{equation}

\item
\underline{
(B)~$\gamma_{\rm b3}<\gamma_{\rm min}<\gamma_{\rm b1}$ 
(i.e., $\nu_{\rm b3}<\nu_{\rm m}<\nu_{\rm b1}$):}\\

In this case, only those electrons with $\gamma>\gamma_{\rm b1}$ can cool.
We have,
\begin{equation}
\nu L_\nu =
\left\{
\begin{array}{ll}
L_0(\nu/\nu_{\rm m})^{(2-p)/2} & ({\rm B1}:~\nu_{\rm b1}<\nu) \\
L_0(\nu_{\rm m}/\nu_{\rm b1})^{1/2}
(\nu/\nu_{\rm m})^{(3-p)/2} & ({\rm B2}:~\nu_{\rm m}<\nu <\nu_{\rm b1})\\
L_0(\nu_{\rm m}/\nu_{\rm b1})^{1/2}
(\nu/\nu_{\rm m})^2 & ({\rm B3}:~\nu <\nu_{\rm m})\\
\end{array}
\right. ~~,
\label{eq:LumiB}
\end{equation}
where $L_0$ is given by Eq.~(\ref{eq:norm2}).
\\

\item
\underline{
(C)~$\gamma_{\rm min}<\gamma_{\rm b3}<\gamma_{\rm b1}$
(i.e., $\nu_{\rm m}<\nu_{\rm b3}<\nu_{\rm b1}$):}\\

In this case, electrons with $\gamma>\gamma_{\rm b1}$ and $\gamma<\gamma_{\rm b3}$
can cool.
The electron spectrum has a break at $\gamma_{\rm b3}$, below which
it is well approximated with the stationary solution,
$\propto \dot{N}(\gamma)t_{\rm Clmb}(\gamma)\propto\gamma^{1-p}$
\citep{sarazin1999}. Hence, we have $L_\nu\propto\nu^{(2-p)/2}$
for $\nu_{\rm m}<\nu <\nu_{\rm b3}$.
Therefore, the luminosity is given by,
\begin{equation}
\nu L_\nu =
\left\{
\begin{array}{ll}
L_0(\nu/\nu_{\rm m})^{(2-p)/2} & ({\rm C1}:~\nu_{\rm b1}<\nu) \\
L_0(\nu_{\rm m}/\nu_{\rm b1})^{1/2}
(\nu/\nu_{\rm m})^{(3-p)/2} & ({\rm C2}:~\nu_{\rm b3}<\nu <\nu_{\rm b1})\\
L_0(\nu_{\rm m}/\nu_{\rm b1})^{1/2}
(\nu_{\rm m}/\nu_{\rm b3})^{1/2}
(\nu/\nu_{\rm m})^{(4-p)/2}
  & ({\rm C3}:~\nu_{\rm m}<\nu <\nu_{\rm b3})\\
L_0(\nu_{\rm m}/\nu_{\rm b1})^{1/2}
(\nu_{\rm m}/\nu_{\rm b3})^{1/2}
(\nu/\nu_{\rm m})^{3/2}
 & ({\rm C4}:~\nu <\nu_{\rm m})\\
\end{array}
\right. ~~,
\label{eq:LumiC}
\end{equation}
where $L_0$ is given by Eq.~(\ref{eq:norm2}).

\end{itemize}

\subsubsection{IC spectrum for  $\nu_{\rm b1}<\nu_{\rm b2}<\nu_{\rm b3}$}

In this case, the electrons suffer significant cooling throughout.
Hence,
an electron with an initial Lorentz factor $\gamma>\gamma_{\rm b2}$
cools down through $\gamma_{\rm b2}$ until which it produces
$P_\nu\propto\nu^{-1/2}$, and further lose its energy to form
$P_\nu\propto\nu^{1/2}$ below $\nu_{\rm b2}$.
The average spectrum from such electrons has a peak at $\nu_{\rm b2}$.
Similarly,
an electron with an initial Lorentz factor $\gamma<\gamma_{\rm b2}$
produces the average spectrum $P_\nu\propto\nu^{1/2}$ for
$\nu<\nu(\gamma)$.
In calculating the net spectrum for power-law electron distribution,
there are two different cases in which 
(D)~$\gamma_{\rm b2}<\gamma_{\rm m}$
and (E)~$\gamma_{\rm m}<\gamma_{\rm b2}$.
Below we describe them in detail.

\begin{itemize}

\item
\underline{
(D) $\gamma_{\rm b2}<\gamma_{\rm min}$ (i.e., $\nu_{\rm b2}<\nu_{\rm m}$):}\\

Electrons with $\gamma>\gamma_{\rm min}$ cools to make a spectrum
$L_\nu\propto\nu^{-p/2}$ for $\nu>\nu_{\rm m}$, while
electrons with $\gamma_{\rm min}$ forms the spectrum below $\nu_{\rm m}$. 
The luminosity is given by,
\begin{equation}
\nu L_\nu =
\left\{
\begin{array}{ll}
L_0(\nu/\nu_{\rm m})^{(2-p)/2} & ({\rm D1}:~\nu_{\rm m}<\nu) \\
L_0(\nu/\nu_{\rm m})^{1/2} & ({\rm D2}:~\nu_{\rm b2}<\nu <\nu_{\rm m})\\
L_0(\nu_{\rm m}/\nu_{\rm b2})(\nu/\nu_{\rm m})^{3/2} & ({\rm D3}:~\nu <\nu_{\rm b2})\\
\end{array}
\right. ~~,
\label{eq:LumiD}
\end{equation}
where $L_0$ is given by Eq.~(\ref{eq:norm2}).
\\

\item
\underline{
(E) $\gamma_{\rm min}<\gamma_{\rm b2}$ (i.e., $\nu_{\rm m}<\nu_{\rm b2}$):}\\

In this case, electron distribution for $\gamma_{\rm min}<\gamma<\gamma_{\rm b2}$
is proportional to $\dot{N}(\gamma)t_{\rm Clmb}(\gamma)\propto\gamma^{1-p}$.
Hence, we have,
\begin{equation}
\nu L_\nu =
\left\{
\begin{array}{ll}
L_0(\nu/\nu_{\rm m})^{(2-p)/2} & ({\rm E1}:~\nu_{\rm b2}<\nu) \\
L_0(\nu_{\rm m}/\nu_{\rm b2})(\nu/\nu_{\rm m})^{(4-p)/2} & ({\rm E2}:~\nu_{\rm m}<\nu <\nu_{\rm b2})\\
L_0(\nu_{\rm m}/\nu_{\rm b2})(\nu/\nu_{\rm m})^{3/2} & ({\rm E3}:~\nu <\nu_{\rm m})\\
\end{array}
\right. ~~,
\label{eq:LumiE}
\end{equation}
where $L_0$ is given by Eq.~(\ref{eq:norm2}).

\end{itemize}

\subsection{Observed surface brightness of IC emission}

The observed surface brightness $S_\nu$ (in units of
erg~s$^{-1}$cm$^{-2}$Hz$^{-1}$str$^{-1}$)
of IC emission from a cluster at redshift $z$ is given by,
\begin{equation}
S_\nu =\frac{(1+z)L_{\nu_{\rm s}}}{4\pi d_L(z)^2\Omega_\nu}~~,
\label{eq:SB}
\end{equation}
where $\nu_{\rm s}=(1+z)\nu$, $d_L(z)$ and $\Omega_{\nu}$ are 
the observing frequency translated into the cluster rest frame,
the luminosity distance to the cluster 
and the solid angle of the extended emission on the sky, respectively.

The {\it effective} value of $\Omega_{\nu}$ can be estimated based on
the observed brightness profile in the sky, which depends on the Lorentz
factor of the electrons, $\gamma_\nu=(\nu/2.05\times10^{11}{\rm Hz})^{1/2}$, 
emitting IC photons with observed frequency $\nu$
[see Eq.~(\ref{eq:chfreq})].
If the total cooling time of such electrons,
$t_{\rm cool}(\gamma_\nu)=[t_{\rm IC/syn}(\gamma_\nu)^{-1}
+t_{\rm Clmb}(\gamma_\nu)^{-1}]^{-1}$,
is much smaller than the dynamical time of the cluster $t_{\rm dyn}$,
then the emission mainly originates from the region around the shock.
The width of the emission region (in the radial direction) is given by the product of
downstream flow velocity and the cooling time,
$\delta\sim(1/3)V_{\rm c}t_{\rm cool}(\gamma_\nu)<r_{\rm vir}$,
where a factor 1/3 applies to  the strong shock limit.
The surface brightness profile has a rim-brightened shape.
Because of projection, the observed apparent scale width, $W$, in the sky
differs from $\delta$. The ratio $W/\delta$ depends on the uncertain radial
profile of the electron distribution downstream of the shock, but
is typically around 3 \citep{bamba2005a}. Writing $W=f_{\rm W}\delta$,
we obtain
$\Omega_\nu\approx2\pi (f_{\rm W}/3)r_{\rm vir}V_{\rm c}t_{\rm cool}(\gamma_\nu)/d_{\rm A}^2$,
where $d_{\rm A}$ is the angular diameter distance to the cluster.
On the other hand, if $t_{\rm cool}(\gamma_\nu)\gg t_{\rm dyn}$, 
the cluster interior is filled with electrons with $\gamma_\nu$.
Taking into account the projection effect, the surface brightness profile is center-filled.
In this case, assuming  the bright emission in the sky originates from inside the radius $r_{\rm vir}/2$,
it occupies a solid angle
$\Omega_\nu\approx(\pi/4)(r_{\rm vir}/ d_{\rm A})^2$.
Connecting both limits, we get
\begin{equation}
\Omega_\nu=\frac{\pi}{4} \left(\frac{r_{\rm vir}}{d_{\rm A}}\right)^2
{\rm min}\left\{1, ~8\frac{f_{\rm W}}{3}\frac{t_{\rm cool}(\gamma_\nu)}{t_{\rm dyn}}
\right\}~~.
\end{equation}


\section{Results}
\label{sec:results}

Based on the derivations in the previous sections, we can now
calculate the observed surface brightness $S_{\nu}$ 
in the SDSS g-band ($\nu=6.3\times10^{14}$Hz)
as a function of 
redshift $z$ and a halo mass $M$.
Other parameters are fixed at the fiducial values, 
$p=2.5$, $r_4=\xi_{{\rm B},-2}=\eta_g=\eta_{e,-5}=1$, 
$\xi_{e,-2}=5$,
$f_{\rm acc}=f_{\rm T}=0.5$ and $f_{\rm W}=3$.
The parameter $\eta_{e}$ is highly uncertain and depends on upstream physical quantities such as
magnetic field and gas temperature \citep[e.g.,][]{matsukiyo2011,guo2014a,guo2014b}. 
It can range from $\sim10^{-7}$ to $\sim10^{-4}$ \citep{kang2012}.
If $\eta_{e}$ is so large, then $\gamma_{\rm min}$ is less than unity 
[see Eq.~(\ref{eq:gminan})], so that our assumption of single power-law spectrum
breaks down. Dependence of $\eta_e$ is easily found as seen in the following.
The spectral index $p$ is also uncertain, however recent study of particle acceleration
at cluster shocks suggest $p\approx2.25$--2.5 \citep{kang2011,kang2013,hong2014,guo2014a}.
In the following, we also consider the cases of $p=2.0$, 2.3 and 3.0, as well as the
fiducial case of $p=2.5$.
The index $p$ is related to the shock compression ratio $r$.
For first-order Fermi acceleration, $p=(r+2)/(r-1)$ in the test-particle limit
\citep{blandford1978,bell1978}.
In this case, $r$ ranges between 2.5 and 4 for $2<p<3$. 
Nevertheless, 
we fix $r_4=(r/4)=1$, which corresponds to the strong shock limit,
because observed surface brightness does not depend on $r$ 
for the parameters of interest.
All the equations in this section do not depend on $r$.

\begin{figure}
\includegraphics[width=80mm]{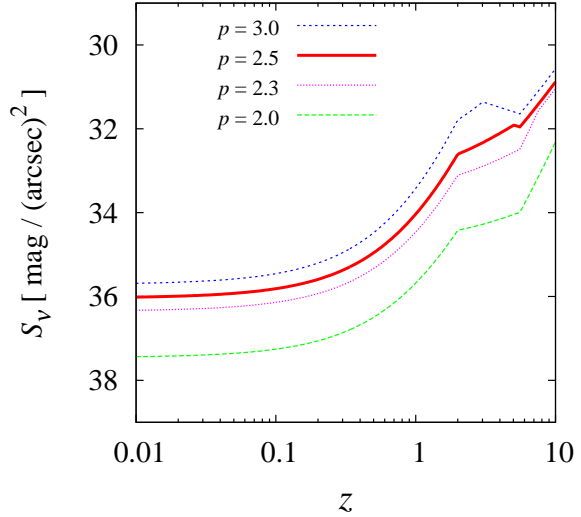}
\caption{
Observed SDSS g-band surface brightness of IC emission from
a cluster with $M_{14}=3$ (in units of mag~arcsec$^{-2}$)  as a function of
cluster redshift $z$. The red line is for fiducial parameters
($p=2.5$, $r_4=\xi_{{\rm B},-2}=\eta_g=\eta_{e,-5}=1$, $\xi_{e,-2}=5$,
$f_{\rm acc}=f_{\rm T}=0.5$ and $f_{\rm W}=3$),
while the green, purple and blue lines are for 
$p=2.0$, 2.3 and 3.0, respectively.
}
\label{fig:brightness}
\end{figure}

\begin{figure}
\includegraphics[width=80mm]{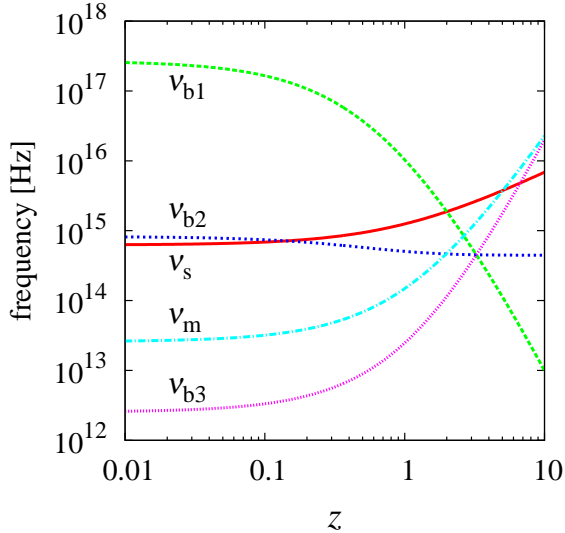}
\caption{
Characteristic frequencies,  as measured in the rest frame of a cluster,
for our fiducial parameter values with a halo mass $M_{14}=3$, as a function of
redshift $z$.
The green, blue, purple and light blue lines indicates
$\nu_{\rm b1}$, $\nu_{\rm b2}$, $\nu_{\rm b3}$ and $\nu_{\rm m}$,
respectively.
The red line represents $\nu_{\rm s}=(1+z)\nu$, which is the observed
frequency blue-shifted to the rest frame of the cluster.
}
\label{fig:freq}
\end{figure}

Figure~\ref{fig:brightness} shows the surface brightness as a function of redshift $z$,
for a fixed halo mass $M_{14}=(M/10^{14}M_{\sun})=3$.
The red line describes the fiducial value of $p=2.5$, while
the others are for different values of $p$ with the other parameters fixed at
their fiducial values.
Interestingly, the surface brightness increases with $z$.
At low redshifts where the effects of cosmological expansion are negligible,
the brightness is almost constant, $\approx36$~mag~arcsec$^{-2}$.
As $z$ increases, the surface brightness becomes larger
by $\approx3$~mag until $z\approx2$.
For this redshift range, 
one can see from Figure~\ref{fig:freq}, that 
$\nu_{\rm b3}<\nu_{\rm m}<\nu_{\rm s}=(1+z)\nu<\nu_{\rm b1}$, 
so that the spectrum is in the regime B2 [see Eq.~(\ref{eq:LumiB})].
Thus, we find,
\begin{eqnarray}
L_{\nu_{\rm s}}&=& L_0\nu_{\rm b1}^{-1/2}\nu_{\rm m}^{(p-2)/2}
\nu_{\rm s}^{(1-p)/2}\nonumber\\
&\propto& f_{\rm acc}f_{\rm T}^{p-1}\xi_e^{p-1}\eta_e^{2-p}M_{14}^{(2p+1)/3}
\nonumber\\
&&~~~\times ~w(z)^{(p-1)/3}(1+z)^{3(p+1)/2}\nu_{\rm s}^{(1-p)/2}~~.
\end{eqnarray}
Since electron cooling is not significant, the solid angle of the emission
is given by $\Omega_\nu\approx(\pi/4)(r_{\rm vir}/ d_{\rm A})^2$, so that,
\begin{eqnarray}
S_{\nu}&\propto& r_{\rm vir}^{-2}(1+z)^{-3}L_{\nu_{\rm s}}\nonumber\\
&\propto& f_{\rm acc}f_{\rm T}^{p-1}\xi_e^{p-1}\eta_e^{2-p}M_{14}^{(2p-1)/3}
\nonumber\\
&&~~~\times ~
w(z)^{(p+1)/3}(1+z)^{p+1}\nu^{(1-p)/2}~~.
\label{eq:Snu1}
\end{eqnarray}
Since $w(z)$ is only weakly dependent on $z$, the surface brightness
$S_\nu$ increases with $z$ following the scaling $(1+z)^{p+1}$.

The regime B2 ends when $\nu_{\rm s}$
becomes larger than $\nu_{\rm b1}$.  This crossing occurs at $z\approx2$ for our
fiducial parameters. Thereafter, the spectrum is in regime B1.
When $z$ further increases, the spectrum enters into regime A1, subsequently
followed by the regime D1.
In these regimes,  the luminosity is given by
the same form [see Eqs.~(\ref{eq:LumiA}), (\ref{eq:LumiB}) and (\ref{eq:LumiD})],
\begin{eqnarray}
L_{\nu_{\rm s}} &=&L_0\nu_{\rm m}^{(p-2)/2}\nu_{\rm s}^{-p/2}\nonumber\\
&\propto& f_{\rm acc}f_{\rm T}^{p-1}\xi_e^{p-1}\eta_e^{2-p}M_{14}^{(2p+1)/3}
\nonumber\\
&&~~~\times ~
w(z)^{(2p+1)/6}(1+z)^{(3p-1)/2}\nu_{\rm s}^{-p/2}~~,
\label{eq:Lnu2}
\end{eqnarray}
and the solid angle is still given by
$\Omega_\nu\approx(\pi/4)(r_{\rm vir}/ d_{\rm A})^2$,
 so that we have,
\begin{eqnarray}
S_\nu &\propto&f_{\rm acc}f_{\rm T}^{p-1}\xi_e^{p-1}\eta_e^{2-p}M_{14}^{(2p-1)/3}
\nonumber\\
&&~~~\times ~
w(z)^{(2p+5)/6}(1+z)^{(2p-3)/2}\nu^{-p/2}~~.
\label{eq:Snu2}
\end{eqnarray}
For our fiducial parameters, Eq.~(\ref{eq:Snu2}) describes the scaling
if $2<z<5.1$.
The flux increases by a factor of about 3  in this redshift range.
Note that the emergence of these regimes originates from  rapid decreasing of 
 $\nu_{\rm b1}\propto(1+z)^{-4}$  for $z>1$.
As a result, the blue-shifted observing frequency $\nu_{\rm s}=(1+z)\nu$ becomes 
larger than any other characteristic frequencies,
$\nu_{\rm m}$ and $\nu_{{\rm b}i}$ ($i=1,2,3$), around $z_{\rm eq}$.
As  $z$ further increases,  $\nu_{\rm m}\propto(1+z)^2$ together with 
$\nu_{\rm b3}\propto(1+z)^4$ become large,
finally exceeding  $\nu_{\rm s}$, so that the spectrum enters the regime D2.
There the electron cooling is so significant that the observed
brightness profile is rim-brightened shape, so that $\Omega_\nu$ is small and
$S_\nu$ shows rapid increase for $z\ga6$.
However, the abundance of clusters at these high redshifts is extremely small 
\citep{barkana2001,watson2013}.


Although the parameter dependence is somewhat complicated, 
one can see that overall behavior is not so different from the fiducial parameter set;
the brightness varies typically by up to $\approx2$--3~mag
if one of parameters is changed with others fixed.
Figure~\ref{fig:brightness} shows  lines for the cases of
$p=2.0$, 2.3 and 3.0 with other parameters fixed as fiducial.
The larger the $p$, the brighter the surface emission.
The dependence on other parameters can be found from
Eqs.~(\ref{eq:Snu1}) and (\ref{eq:Snu2}).


\begin{figure}
\includegraphics[width=80mm]{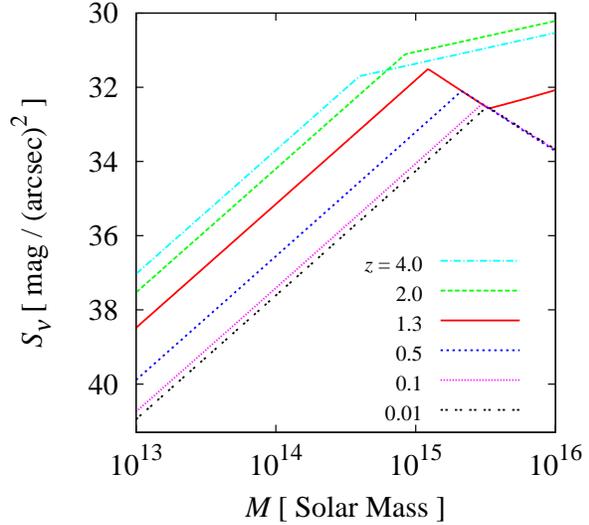}
\caption{
Observed SDSS g-band surface brightness (in units of mag~arcsec$^{-2}$) as a function of
a halo mass $M$ for our fiducial parameters
($p=2.5$, $r_4=\xi_{{\rm B},-2}=\eta_g=\eta_{e,-5}=1$, $\xi_{e,-2}=5$,
$f_{\rm acc}=f_{\rm T}=0.5$ and $f_{\rm W}=3$).
Different lines correspond to different redshifts $z$ of the
cluster.
}
\label{fig:brightness2}
\end{figure}


Figure~\ref{fig:brightness2} depicts the surface brightness as a function of
a halo mass $M$ for fixed cluster redshifts assuming the fiducial parameter values. 
As seen in Fig.~\ref{fig:brightness},
the redshift dependence is small when $z\la0.1$.
There, the surface brightness peaks at $M_{14}\approx30$, below which
the spectrum is in regime B2 for $0.55\la M_{14}\la30$
and in regime C2 for $M_{14}\la0.55$.
Functional forms of the brightness for B2 and C2 are given by Eq.~(\ref{eq:Snu1}),
with $S_\nu\propto M_{14}^{(2p-1)/3}$.
For $M_{14}\ga30$ the spectrum is in regime B3, at which we have,
\begin{equation}
S_\nu\ \propto
f_{\rm acc}f_{\rm T}^{-2}\xi_e^{-2}\eta_e^{3}M_{14}^{-1}w(z)\nu~~,
\label{eq:Snu3}
\end{equation}
so that the brightness decreases with $M$.
For $z=0.5$, the peak shifts to lower masses, and larger values.
Note that Eq.~(\ref{eq:Snu3}) implies that the brightness in regime B3 very weakly
depends on $z$.
The behavior for $z=1.3$ is the same as for lower redshifts as long as
$M_{14}<33$, above which, however, the spectrum is in the regime A3 and
the surface brightness again increases with $M$ and $z$
 according to the scaling,
\begin{equation}
S_\nu\propto
f_{\rm acc}\eta_e M_{14}^{1/3}w(z)^{1/3}(1+z)^7\nu~~.
\label{eq:Snu4}
\end{equation}
For higher redshifts $z=2.0$ ($z=4.0$), the spectrum enters
regimes C2, B2 and B1 (E1 and D1) in turn towards higher $M$.
The brightness
breaks at $M_{14}=8.5$ and 3.9 for $z=2.0$ and 4.0,
respectively. After the break, the brightness still increases with $M$
according to the scaling,
\begin{equation}
S_\nu\propto
f_{\rm acc}\eta_e M_{14}^{1/3}w(z)^{7/6}(1+z)^{-1/2}\nu^{-1/2}~~.
\label{eq:Snu5}
\end{equation}
In summary, the surface brightness follows simple
scaling behavior, $S_\nu\propto M^{(2p-1)/3}$,
for the typical range of expected parameters.


\section{Discussion}
\label{sec:summary}

Using a simple analytical model, 
we have calculated IC emission in the SDSS g-band from relativistic electrons
accelerated in galaxy clusters, taking into account the effects of 
Coulomb, synchrotron, and IC energy loss of the emitting electrons.
For our fiducial parameters, at  $z\la2$ and $M\la10^{15}M_{\sun}$,
the spectrum is in the regime B2 or C2, in which 
$S_\nu\propto M^{(2p-1)/3}(1+z)^{p+1}\nu^{(1-p)/2}$, where
$p$ is the power-law index of electron distribution (see Fig.~\ref{fig:brightness2}). 
If the value of $p$ is inferred from radio synchrotron emission,
one can predict the spectral index of the optical IC emission.
In this paper,
we have not taken into account the possibility of
reacceleration in the downstream turbulence,
which enhances the abundance of relativistic electrons 
\citep{schlickeiser1987,brunetti2001,petrosian2001,fujita2003}, 
resulting in brighter IC emission.
We also assume a spherical, virialized shock as a result of smooth accretion.
If the cluster is not dynamically relaxed and is still being formed, e.g. as a result 
of a merger of two sub-components, then the shock structure would be different 
and turbulent energy would be enhanced.
In this case,  we expect brighter IC emission due to more efficient reacceleration. 
Hence,  our estimate for the IC flux and surface brightness should be
regarded as a conservative lower limit.

It would be natural to assume that electron acceleration occurs
at the virial shocks.
In merging clusters, radio relics have been
detected as possible signatures of the merger shocks, although clear
evidence for it had not been identified as of yet.
Observations at other wavelengths would be helpful for shock identification.
Proton acceleration at the virial shocks could also occur, although we have only a few 
observational implications of it \citep[e.g.,][]{fujita2013}.  
Observations of optical IC emission may have the advantage of enabling 
the identification of shocks,
because the angular resolution of the optical telescopes is generally much better than 
hard X-ray or gamma-ray telescopes.
According to  the theory of  diffusive shock acceleration, electrons responsible for
the optical IC emission cannot penetrate far upstream relative to the shock front.
Hence,  the IC emissivity is expected to have a steep rise across the shock front.
The detection of such a sharp jump in the IC emission would flag  the position 
of the shock front, and provide  evidence for  particle acceleration there. 
In our spherical model, however, IC-emitting electrons are not rapidly cooling
 in most cases, resulting in center-filled shape of the brightness profile
on the sky.
Thus, the profile may not have a sharp rise in projection.
Possible exceptions might be expected for more realistic, non-spherical cases,
where sharp feature could be found  on the sky
at the location of shocks.
Detailed morphological studies based on numerical simulations, 
like the case of radio relics \citep[e.g.,][]{skillman2013,hong2015},
are needed in this case and go beyond the scope of this paper.

Our model predicts the surface brightness of IC emission from
several massive clusters, as summarized in Table~1.
Larger fluxes are expected for
 larger $M$ and  $z$ if the spectrum is in regime B2 or C2
 [see Eq.~(\ref{eq:Snu1})].
The predicted brightness ranges between $\approx32$ and 35~mag~arcsec$^{-2}$.
Such ultralow surface brightnesses should
 be detectable  with the recently developed 
Dragonfly Telephoto Array \citep{abraham2014}. 
More detailed prospects for the observation of individual clusters will be
given elsewhere.


\begin{table*}
\label{table:brightness}
\centering
\begin{minipage}{140mm}
\caption{Predicted surface brightness in SDSS g-band for specific clusters.}
\begin{tabular}{@{}llrcccc@{}}
\hline
%
Name & $z$ 
& $M$\footnote{Values of $M_{200}$ are taken from references.} 
& \multicolumn{3}{c}{Surface brightness\footnote
{For fiducial parameters other than $p$.
Spectral regime is also shown in parentheses.}} & Reference \\
& &  [$10^{14}M_{\sun}$] & \multicolumn{3}{c}{[mag~arcsec$^{-2}$]} & \\
& & & $p=2.0$ & $p=2.5$ & $p=3.0$ & \\
\hline
\hline
IDCS J1426.5$+$3508 & 1.75 & 5.3 & 34.1 (C2) & 32.1 (B2) & 31.2  (B3)
& \citet{stanford2012}  \\
SPT-CL J2106$-$5844 & 1.132 & 9.8 & 34.2 (B2) & 32.1 (B2) & 31.8  (B3)
& \citet{williamson2011} \\
ACT-CL J0102$-$4915 & 0.870 & 22.3 & 33.7 (B2) &32.1 (B3)& 32.7 (B3)
& \citet{menanteau2012} \\
SPT-CL J2344$-$4243 & 0.596 & 25.0 & 34.0 (B2) & 32.3 (B3) & 32.8 (B3)
& \citet{mcdonald2012} \\
MS~1054$-$0321 & 0.83 & 12 & 34.4 (B2) & 32.3 (B2) & 32.0  (B3)
& \citet{jee2005} \\
SPT-CL J0658$-$5556 & 0.296 & 31.2 & 34.3 (B2) & 32.5 (B3) & 33.1 (B3)
& \citet{williamson2011} \\
XDCP J0044.0$-$2033 & 1.579 & 4.4 & 34.5 (C2) & 32.6 (B2) & 31.7  (B2)
& \citet{tozzi2015} \\
SPT-CL J2337$-$5942 & 0.775 & 10.5 & 34.7 (B2) & 32.6 (B2) & 31.9  (B3)
& \citet{williamson2011} \\
Coma & 0.0232 & 27.8 & 35.0 (B2) & 32.7 (B2) & 32.9  (B3)
& \citet{kubo2007} \\
MACS~J1206.2$-$0847 & 0.44 & 14.1 & 34.9 (B2) & 32.8 (B2) & 32.2  (B3)
& \citet{presotto2014} \\
Abell 2390 & 0.228 & 18 & 35.1 (B2) & 32.9 (B2) & 32.5  (B3)
& \citet{carlberg1996} \\
XLSSU J021744.1$-$034536 & 1.91 & $\approx2$\footnote{
Since only $M_{500}$ is given in \citet{mantz2014}, we estimate $M_{200}$ assuming
isothermal density distribution as $M_{200}=1.58M_{500}$.
} & 35.0 (C2) & 33.3 (B2) & 32.6 (B2)
& \citet{mantz2014}  \\
Abell 2744 & 0.3064 & 70 & 33.4 (B2) & 33.3 (B3) & 33.9  (B3)
& \citet{montes2014} \\
\hline
\end{tabular}
\end{minipage}
\end{table*}


Our model also predicts the color of the IC emission,
which is bluer than starlight.
If we assume a power-law form of the IC emission, as
$S_\nu\propto\nu^{-\alpha}$, then the color
$g-r$ is calculated as
\begin{equation}
g-r = \frac{5}{2}\alpha\log_{10}(\nu_g/\nu_r) = 0.29\alpha~~,
\end{equation}
where $\nu_g=6.3\times10^{14}$Hz and $\nu_r=4.8\times10^{14}$Hz
are SDSS g-band and r-band frequencies, respectively.
In the case of spectral regime B2 or C2 (i.e., $\alpha=(p-1)/2$), we obtain 
$g-r=0.19$ for $p=2.3$.
If the spectral regime is in B3 (i.e., $\alpha=-1$), then $g-r=-0.29$.
These values are distinguishable from stellar components, such as 
diffuse faint emission of brightest cluster galaxies (BCG), satellite galaxies,
and intracluster light (ICL), which typically shows $g-r\ga0.7$
\citep[e.g.,][]{montes2014}.

\begin{figure}
\includegraphics[width=80mm]{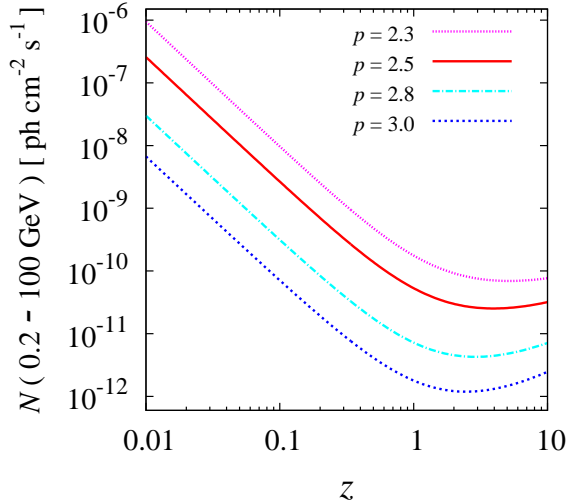}
\caption{
Observed IC gamma-ray flux $N(0.2-100~{\rm GeV})$, in units of
photons~cm$^{-2}$~s$^{-1}$, as  a function of redshift $z$
for most massive ($M_{14}=30$) clusters.
The red line is for fiducial parameters
($p=2.5$, $r_4=\xi_{{\rm B},-2}=\eta_g=\eta_{e,-5}=1$, $\xi_{e,-2}=5$,
$f_{\rm acc}=f_{\rm T}=0.5$ and $f_{\rm W}=3$),
while the purple, light blue and blue lines are for 
$p=2.3$, 2.8 and 3.0, respectively.
Mass dependence of the gamma-ray flux is given by Eq.~(\ref{eq:Lnu2}),
as $N(0.2-100~{\rm GeV})\propto M_{14}^{(2p+1)/3}$,
because observed frequency blue-shifted to the rest frame,
$\nu_{\rm s}$, is always larger than any other characteristic frequencies,
$\nu_{\rm b1}$, $\nu_{\rm b2}$, $\nu_{\rm b3}$ and $\nu_{\rm m}$.
Observed upper limits for specific clusters lie 
$N(0.2-100~{\rm GeV})\la10^{-9}$--$10^{-8}$photons~cm$^{-2}$s$^{-1}$
\citep{ackermann2010}.
}
\label{fig:gamma}
\end{figure}

\begin{figure}
\includegraphics[width=80mm]{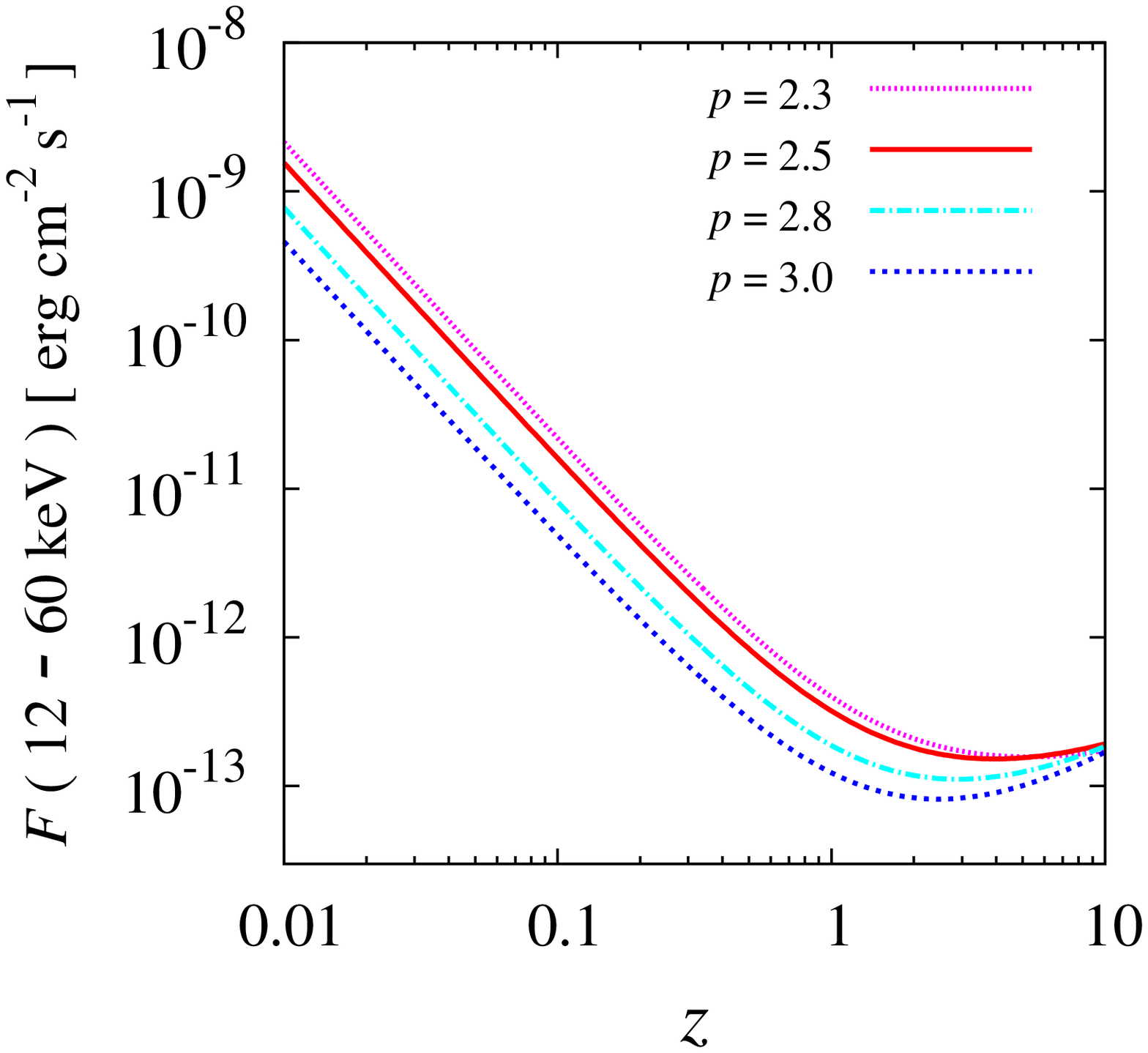}
\caption{
Observed IC X-ray flux $F(12-60~{\rm keV})$, in units of
erg~cm$^{-2}$~s$^{-1}$, as  a function of redshift $z$
for most massive ($M_{14}=30$) clusters.
The red line is for fiducial parameters
($p=2.5$, $r_4=\xi_{{\rm B},-2}=\eta_g=\eta_{e,-5}=1$, $\xi_{e,-2}=5$,
$f_{\rm acc}=f_{\rm T}=0.5$ and $f_{\rm W}=3$),
while the purple, light blue and blue lines are for 
$p=2.3$, 2.8 and 3.0, respectively.
Mass dependence of the gamma-ray flux is given by Eq.~(\ref{eq:Lnu2}),
as $F(12-60~{\rm keV})\propto M_{14}^{(2p+1)/3}$,
because observed frequency blue-shifted to the rest frame,
$\nu_{\rm s}$, is always larger than any other characteristic frequencies,
$\nu_{\rm b1}$, $\nu_{\rm b2}$, $\nu_{\rm b3}$ and $\nu_{\rm m}$.
Observed upper limits for specific clusters lie 
$F(12-60~{\rm keV})\la10^{-11}$--$10^{-10}$erg~cm$^{-2}$s$^{-1}$
\citep{ota2014}.
}
\label{fig:X}
\end{figure}

Gamma-ray and hard X-ray observations have given upper limits,
which constrain our model parameters.
Figures \ref{fig:gamma} and \ref{fig:X} show that
 for our fiducial parameters (namely $p=2.5$), both IC gamma-ray and
X-ray fluxes for nearby ($z\ll0.1$), massive ($M_{14}=30$) clusters exceed current
observational upper limits, which are
$N(0.2-100~{\rm GeV})\la10^{-9}$--$10^{-8}$photons~cm$^{-2}$s$^{-1}$
for gamma-rays \citep{ackermann2010} and
$F(12-60~{\rm keV})\la10^{-11}$--$10^{-10}$erg~cm$^{-2}$s$^{-1}$
for X-rays \citep{ota2014}.
Although the case with $p\la2.5$ is unlikely for such nearby massive clusters,
our model predictions can be lower than these current observational
upper limits for less massive ($M_{14}\la10$), higher redshift ($z\ga0.1$) 
or steeper electron index ($p\ga2.8$)\footnote{
Some X-ray observations for specific clusters 
have provided tight upper limits, such as
$F(12-60~{\rm keV})<5\times10^{-12}$erg~cm$^{-2}$s$^{-1}$
\citep{nishino2010}
for Perseus ($z=0.018$ and $M_{200}=1.2\times10^{15}M_{\sun}$),
while our model overpredicts X-ray flux by a factor of 3 larger even for $p=3.0$.
In such cases, smaller value of $\xi_e$ ($\la0.01$) may be required.
}.
Note that we conservatively adopt $p=2.5$ in this paper, but that higher values of $p$ 
as implied by the gamma-ray and X-ray limits for low redshift clusters 
would make our predicted optical IC flux somewhat higher (see Fig.~\ref{fig:brightness}). 
Another independent way to have lower gamma-ray flux is to adopt
unusually large $\eta_g\gg10^2$. In this case, $\gamma_{\rm max}$ is small enough
to give spectral cutoff of IC emission at energy below the {\it Fermi} band.
%


The IC emission from the virial shock of clusters could be enhanced due to 
relativistic electrons produced in supernovae that gradually diffuse out of the cluster. 
The latter contribution depends on the star formation history and the diffusion time 
of relativistic electrons within the cluster. 
The diffusion time of electrons depends on the unknown configuration of magnetic fields. 
If the fields are radially aligned in the outer envelope of clusters 
\citep[as expected from radial infall or the magneto-thermal instability; see e.g.,][]{parrish2012},
then the diffusion time there would be of order the light crossing time of the outer parts 
of the cluster, i.e. millions of years. 
The {\it Fermi} satellite has placed tight limits on this contribution in cluster
cores based on the lack of gamma-ray emission at 0.2--100 GeV there
\citep[][see also Vazza et al. 2015]{ackermann2010,ackermann2014}.

\begin{figure}
\includegraphics[width=80mm]{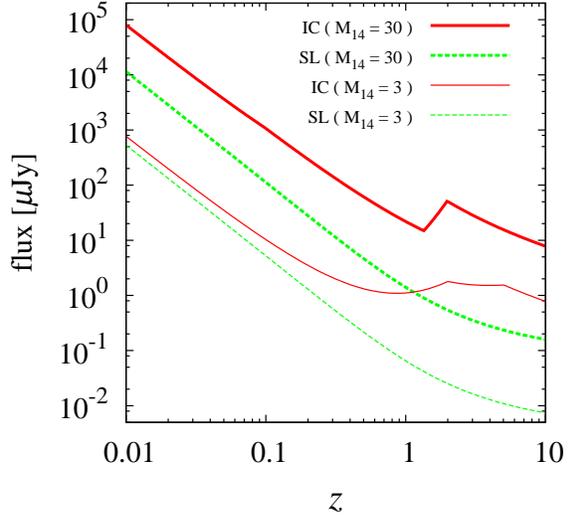}
\caption{
Observed SDSS g-band flux (in units of $\mu$Jy) of
IC emission (red lines) and the Thomson scattered starlight (green lines) 
as a function of redshift $z$.
Thick and thin lines are for a halo mass $M_{14}=30$ and $M_{14}=3$,
respectively.
In calculating the IC flux, we adopt fiducial parameters
($p=2.5$, $r_4=\xi_{{\rm B},-2}=\eta_g=\eta_{e,-5}=1$, $\xi_{e,-2}=5$,
$f_{\rm acc}=f_{\rm T}=0.5$ and $f_{\rm W}=3$).
}
\label{fig:flux}
\end{figure}

Diffuse optical emission from Thomson scattering of starlight could be comparable 
to the IC emission.
We roughly estimate the flux of the scattered light emission as,
\begin{equation}
\nu F_\nu^{\rm (sc)} \sim\frac{L_*\tau_{\rm T}}{4\pi d_L(z)^2}~~,
\end{equation}
where $L_*$ is the bolometric stellar luminosity and 
$\tau_{\rm T}$ is the optical depth for the Thomson scattering.
The typical mass-to-light ratio within the virial
radius $r_{\rm vir}$ is given by $M/L\approx2.5\times10^2M_{\sun} /L_{\sun}$
\citep[e.g.,][]{sheldon2009,holland2015},
so that $L_*\approx1.6\times10^{45}M_{14}$~erg~s$^{-1}$.
The optical depth $\tau_{\rm T}$  can be estimated as,
\begin{eqnarray}
\tau_{\rm T} &\sim& n_e\sigma_{\rm T}r_{\rm vir}\nonumber \\
&\sim&3.3\times10^{-4}r_4f_{\rm acc}M_{14}^{1/3}w(z)^{2/3}(1+z)^2~~.
\end{eqnarray}
Note that this estimate may be upper limit since we implicitly assume that most of the
energy of the starlight is contained in the SDSS g-band in the observer frame.
In Figure~\ref{fig:flux}, we show $F_\nu^{\rm (sc)}$ as a function of redshift $z$ (green lines),
comparing with the IC flux (red lines).
One can see that the IC emission dominates if $M_{14}\ga3$.
Since the spectrum of the scattered light is very different from that of the IC emission,
color measurements can be used to distinguish between them.


The BCG and ICL could also constitute a diffuse background.
The ICL emission has been calculated based on the cosmological simulation
\citep[e.g.,][]{rudick2011,laporte2013,cui2014}.
According to the recent result by \citet{cui2014},
the ICL brightness could be around 30~mag~arcsec$^{-2}$ at $r_{\rm vir}$ for
the most massive clusters.
On the other hand,
many observations measure the  brightness profile of the BCG+ICL
 only in the interior of clusters due to its faintness
\citep[e.g.,][]{presotto2014,demaio2015}. 
Extrapolating linearly the observed
radial profile of BCG+ICL of a cluster MACS~J1206.2$-$0847 \citep{presotto2014}
to its virial radius of $r_{\rm vir}\sim2.3$~Mpc yields emission that
is dimmer than $\sim35$~mag~arcsec$^{-2}$, so that
IC emission is  brighter for this cluster (see Table~1).
Even if BCG+ICL has comparable brightness to the IC emission,
 their color difference can be used to separate them from each other.

IC emission also lies in infrared bands; however, the flux is rather small because
typically $\nu_{\rm s}<\nu_{\rm m}$ (i.e., it is in regime B3).
Furthermore, mid and far-infrared bands may be  dominated by dust
emission \citep{yamada2005,kitayama2009}.

Finally, we remark on IC emission as a possible background emission 
 for other purposes. 
For example, as already discussed above, 
the BCG+ICL brightness around the virial radius $r_{\rm vir}$ could be comparable
to the IC emission.
The surface brightness  of dwarf galaxies \citep[e.g.,][]{herrmann2013}
is also in
 some cases  less than 32 mag~arcsec$^{-2}$ at the 
periphery of the galaxies.
The IC emission could confuse or disguise these different emissions
from member galaxies in massive clusters.


\section*{Acknowledgments}

We thank 
Roberto~Abraham,
Yutaka~Fujita,  
Hyesung~Kang,
Yutaka~Ohira, 
Kouji~Ohta,
Dongsu~Ryu,
Lorenzo~Sironi,
Shuta~Tanaka,
Makoto~Uemura
and
Pieter~van~Dokkum,
for valuable comments. 
We also thank the anonymous referee for valuable comments to improve the paper.
This work was supported in part by grant-in-aid 
from the Ministry of Education, Culture, Sports, 
Science, and Technology (MEXT) of Japan, 
No.~15K05088 (R.~Y.) and 
NSF grant AST-1312034 (A.~L.).
R.~Y. also thank ISSI (Bern) for support of the team 
``Physics of the Injection of Particle Acceleration at 
Astrophysical, Heliospheric, and Laboratory Collisionless Shocks''. 






\label{lastpage}

\begin{thebibliography}{99}
%

\bibitem[Abraham \& van Dokkum(2014)]{abraham2014} 
Abraham, R.~G., \& van Dokkum, P.~G.\ 2014, PASP, 126, 55 

\bibitem[Ackermann et al.(2010)]{ackermann2010} 
Ackermann, M., Ajello, M., Allafort, A., et al.\ 2010, ApJL, 717, L71 

\bibitem[Ackermann et al.(2014)]{ackermann2014} 
Ackermann, M., Ajello, M., Albert, A., et al.\ 2014, ApJ, 787, 18 

\bibitem[Ade et al.(2015)]{planck2015} 
Ade, P. A. R. et al. (Planck Collaboration) 2015, arXiv:1502.01589

\bibitem[\protect\citeauthoryear{Bamba et al.}{2003}]{bamba2003}
Bamba, A. et al. 2003, ApJ, 589, 827 

\bibitem[\protect\citeauthoryear{Bamba et al.}{2005a}]{bamba2005a}
Bamba, A. et al. 2005a,  ApJ, 621, 793
%
\bibitem[\protect\citeauthoryear{Bamba et al.}{2005b}]{bamba2005b}
Bamba, A. et al. 2005b,  ApJ, 632, 294

\bibitem[Bartels et al.(2015)]{bartels2015} 
Bartels, R., Zandanel, F., \& Ando, S.\ 2015, A\&A in press (arXiv:1501.06940) 

\bibitem[\protect\citeauthoryear{Bell}{1978}]{bell1978}
Bell, A. R. 1978, MNRAS, 182, 147

\bibitem[Barkana \& Loeb(2001)]{barkana2001} 
Barkana, R., \& Loeb, A.\ 2001, Phys. Rep., 349, 125 


\bibitem[\protect\citeauthoryear{Blandford \& Ostriker}{1978}]{blandford1978}
Blandford, R. D., \& Ostriker, J. P. 1978, ApJ, 221, L29

\bibitem[Blasi \& Colafrancesco(1999)]{blasi1999} 
Blasi, P., \& Colafrancesco, S.\ 1999, Astroparticle Physics, 12, 169 


\bibitem[Blumenthal \& Gould(1970)]{blumenthal1970} 
Blumenthal, G.~R., \& Gould, R.~J.\ 1970, Reviews of Modern Physics, 42, 237 

\bibitem[Brunetti et al.(2001)]{brunetti2001} 
Brunetti, G., Setti, G., Feretti, L., \& Giovannini, G.\ 2001, MNRAS, 320, 365 

\bibitem[Brunetti \& Jones(2014)]{brunetti2014} 
Brunetti, G., \& Jones, T.~W.\ 2014, Int. J. of Mod. Phys. D, 23, 1430007 



\bibitem[Bryan \& Norman(1998)]{bryan1998} 
Bryan, G.~L., \& Norman, M.~L.\ 1998, ApJ, 495, 80 

\bibitem[Carlberg et al.(1996)]{carlberg1996} 
Carlberg, R.~G., Yee, H.~K.~C., Ellingson, E., et al.\ 1996, ApJ, 462, 32 

\bibitem[Cui et al.(2014)]{cui2014} 
Cui, W., Murante, G., Monaco, P., et al.\ 2014, MNRAS, 437, 816 

\bibitem[DeMaio et al.(2015)]{demaio2015} 
DeMaio, T., Gonzalez, A.~H., Zabludoff, A., Zaritsky, D., \& Brada{\v c}, M.\ 2015, MNRAS, 448, 1162 

\bibitem[\protect\citeauthoryear{Drury}{1983}]{drury83}
Drury, L. O'C., 1983, Rep. Prog. Phys., 46, 973

\bibitem[Fixsen(2009)]{fixsen2009} 
Fixsen, D. J. 2009, ApJ, 707, 916

\bibitem[Fujita \& Kato(2005)]{fujita2005} 
Fujita, Y., \& Kato, T.~N.\ 2005, MNRAS, 364, 247 

\bibitem[Fujita et al.(2013)]{fujita2013} 
Fujita, Y., Ohira, Y., \& Yamazaki, R.\ 2013, ApJL, 767, L4 

\bibitem[Fujita \& Sarazin(2001)]{fujita2001} 
Fujita, Y., \& Sarazin, C.~L.\ 2001, ApJ, 563, 660 

\bibitem[Fujita et al.(2003)]{fujita2003} 
Fujita, Y., Takizawa, M., \& Sarazin, C.~L.\ 2003, ApJ, 584, 190 

\bibitem[Gastaldello et al.(2015)]{gastaldello2015} 
Gastaldello, F., Wik, D.~R., Molendi, S., et al.\ 2015, ApJ, 800, 139 

\bibitem[Guo et al.(2014a)]{guo2014a} 
Guo, X., Sironi, L., \& Narayan, R.\ 2014a, ApJ, 794, 153 

\bibitem[Guo et al.(2014b)]{guo2014b} 
Guo, X., Sironi, L., \& Narayan, R.\ 2014b, ApJ, 797, 47 

\bibitem[Herrmann et al.(2013)]{herrmann2013} 
Herrmann, K.~A., Hunter, D.~A., \& Elmegreen, B.~G.\ 2013, AJ, 146, 104 

\bibitem[Holland et al.(2015)]{holland2015} 
Holland, J.~G., B{\"o}hringer, H., Chon, G., \& Pierini, D.\ 2015, MNRAS, 448, 2644 

\bibitem[Hong et al.(2014)]{hong2014} 
Hong, S.~E., Ryu, D., Kang, H., \& Cen, R.\ 2014, ApJ, 785, 133 

\bibitem[Hong et al.(2015)]{hong2015} 
Hong, S.~E., Kang, H., \& Ryu, D.\ 2015, arXiv:1504.03102 

\bibitem[Inoue et al.(2005)]{inoue2005} 
Inoue, S., Aharonian, F.~A., \& Sugiyama, N.\ 2005, ApJL, 628, L9 

\bibitem[Jee et al.(2005)]{jee2005} 
Jee, M.~J., White, R.~L., Ford, H.~C., et al.\ 2005, ApJ, 634, 813 

\bibitem[Kang \& Ryu(2011)]{kang2011} 
Kang, H., \& Ryu, D.\ 2011, ApJ, 734, 18 

\bibitem[Kang et al.(2012)]{kang2012} 
Kang, H., Ryu, D., \& Jones, T.~W.\ 2012, ApJ, 756, 97 

\bibitem[Kang \& Ryu(2013)]{kang2013} 
Kang, H., \& Ryu, D.\ 2013, ApJ, 764, 95 

\bibitem[Keshet et al.(2003)]{keshet2003} 
Keshet, U., Waxman, E., Loeb, A., Springel, V., \& Hernquist, L.\ 2003, ApJ, 585, 128 

\bibitem[Keshet et al.(2004a)]{keshet2004a} 
Keshet, U., Waxman, E., \& Loeb, A.\ 2004a, JCAP, 4, 006 

\bibitem[Keshet et al.(2004b)]{keshet2004b} 
Keshet, U., Waxman, E., \& Loeb, A.\ 2004b, ApJ, 617, 281 

\bibitem[Keshet et al.(2012)]{keshet2012} 
Keshet, U., Kushnir, D., Loeb, A., \& Waxman, E.\ 2012, arXiv:1210.1574 

\bibitem[Kitayama et al.(2009)]{kitayama2009} 
Kitayama, T., Ito, Y., Okada, Y., et al.\ 2009, ApJ, 695, 1191 

\bibitem[Kubo et al.(2007)]{kubo2007} 
Kubo, J.~M., Stebbins, A., Annis, J., et al.\ 2007, ApJ, 671, 1466 

\bibitem[Kushnir \& Waxman(2009)]{kushnir2009} 
Kushnir, D., \& Waxman, E.\ 2009, JCAP, 8, 002 

\bibitem[Kushnir \& Waxman(2010)]{kushnir2010} 
Kushnir, D., \& Waxman, E.\ 2010, JCAP, 2, 025 

\bibitem[Laporte et al.(2013)]{laporte2013} 
Laporte, C.~F.~P., White, S.~D.~M., Naab, T., \& Gao, L.\ 2013, MNRAS, 435, 901 

\bibitem[Lau et al.(2015)]{lau2014} 
Lau, E. T., Nagai, D., Avestruz, C. et al. 2015, ApJ, 806, 68

\bibitem[Loeb \& Waxman(2000)]{loeb2000} 
Loeb, A., \& Waxman, E.\ 2000, Nature, 405, 156 


\bibitem[Matsukiyo et al.(2011)]{matsukiyo2011} 
Matsukiyo, S., Ohira, Y., Yamazaki, R., \& Umeda, T.\ 2011, ApJ, 742, 47 

\bibitem[McDonald et al.(2012)]{mcdonald2012} 
McDonald, M., Bayliss, M., Benson, B.~A., et al.\ 2012, Nature, 488, 349 

\bibitem[Mantz et al.(2014)]{mantz2014} 
Mantz, A.~B., Abdulla, Z., Carlstrom, J.~E., et al.\ 2014, ApJ, 794, 157 

\bibitem[Menanteau et al.(2012)]{menanteau2012} 
Menanteau, F., Hughes, J.~P., Sif{\'o}n, C., et al.\ 2012, ApJ, 748, 7 

\bibitem[Miniati(2002)]{miniati2002} 
Miniati, F.\ 2002, MNRAS, 337, 199 

\bibitem[Miniati(2003)]{miniati2003} 
Miniati, F.\ 2003, MNRAS, 342, 1009 


\bibitem[Miniati(2015)]{miniati2015} 
Miniati, F.\ 2015, ApJ, 800, 60 

\bibitem[Montes \& Trujillo(2014)]{montes2014} 
Montes, M., \& Trujillo, I.\ 2014, ApJ, 794, 137 

\bibitem[Nelson et al.(2015)]{nelson2015} 
Nelson, D., Genel, S., Pillepich, A., et al.\ 2015, arXiv:1503.02665 

\bibitem[Nishino et al.(2010)]{nishino2010} 
Nishino, S., Fukazawa, Y., Hayashi, K., Nakazawa, K., \& Tanaka, T.\ 2010, PASJ, 62, 9 

\bibitem[Ota et al.(2014)]{ota2014} 
Ota, N., Nagayoshi, K., Pratt, G.~W., et al.\ 2014, A\&A, 562, A60 

\bibitem[Parrish et al.(2012)]{parrish2012} 
Parrish, I.~J., McCourt, M., Quataert, E., \& Sharma, P.\ 2012, MNRAS, 419, L29 

\bibitem[Patej \& Loeb(2015)]{patej2015} 
Patej, A., \& Loeb, A.\ 2015, ApJL, 798, L20 

\bibitem[Petrosian(2001)]{petrosian2001} 
Petrosian, V.\ 2001, ApJ, 557, 560 

\bibitem[Petrosian et al.(2008)]{petrosian2008} 
Petrosian, V., Bykov, A., \& Rephaeli, Y.\ 2008, Sp. Sci. Rev., 134, 191 

\bibitem[Presotto et al.(2014)]{presotto2014} 
Presotto, V., Girardi, M., Nonino, M., et al.\ 2014, A\&A, 565, A126 



\bibitem[Rephaeli et al.(2008)]{rephaeli2008} 
Rephaeli, Y., Nevalainen, J., Ohashi, T., \& Bykov, A.~M.\ 2008, Sp. Sci. Rev., 134, 71 

\bibitem[Rudick et al.(2011)]{rudick2011} 
Rudick, C.~S., Mihos, J.~C., \& McBride, C.~K.\ 2011, ApJ, 732, 48 

\bibitem[Ryu et al.(2003)]{ryu2003} 
Ryu, D., Kang, H., Hallman, E., \& Jones, T.~W.\ 2003, ApJ, 593, 599 

\bibitem[Ryu et al.(2008)]{ryu2008} 
Ryu, D., Kang, H., Cho, J., \& Das, S.\ 2008, Science, 320, 909 

\bibitem[Sarazin(1999)]{sarazin1999} 
Sarazin, C.~L.\ 1999, ApJ, 520, 529 

\bibitem[Sari, Piran, \& Narayan(1998)]{sari1998}
Sari, R., Piran, T., \& Narayan, R. 1998, ApJL, 497, L17 

\bibitem[Schaal \& Springel(2015)]{schaal2015} 
Schaal, K., \& Springel, V.\ 2015, MNRAS, 446, 3992 

\bibitem[Schlickeiser et al.(1987)]{schlickeiser1987} 
Schlickeiser, R., Sievers, A., \& Thiemann, H.\ 1987, A\&A, 182, 21 


\bibitem[Sheldon et al.(2009)]{sheldon2009} 
Sheldon, E.~S., Johnston, D.~E., Masjedi, M., et al.\ 2009, ApJ, 703, 2232 

\bibitem[Skillman et al.(2013)]{skillman2013} 
Skillman, S.~W., Xu, H., Hallman, E.~J., et al.\ 2013, ApJ, 765, 21 

\bibitem[Stanford et al.(2012)]{stanford2012} 
Stanford, S.~A., Brodwin, M., Gonzalez, A.~H., et al.\ 2012, ApJ, 753, 164 

\bibitem[Takizawa \& Naito(2000)]{takizawa2000} 
Takizawa, M., \& Naito, T.\ 2000, ApJ, 535, 586 

\bibitem[Takizawa(2002)]{takizawa2002} 
Takizawa, M.\ 2002, PASJ, 54, 363 

\bibitem[Takizawa(2008)]{takizawa2008} 
Takizawa, M.\ 2008, ApJ, 687, 951 


\bibitem[Totani \& Kitayama(2000)]{totani2000} 
Totani, T., \& Kitayama, T.\ 2000, ApJ, 545, 572 



\bibitem[Tozzi et al.(2015)]{tozzi2015} 
Tozzi, P., Santos, J.~S., Jee, M.~J., et al.\ 2015, ApJ, 799, 93 

\bibitem[van Dokkum et al.(2014)]{vanDokkum2014} 
van Dokkum, P.~G., Abraham, R., \& Merritt, A.\ 2014, ApJL, 782, L24 

\bibitem[Vazza et al.(2015)]{vazza2015} 
Vazza, F., Eckert, D., Brueggen, M., \& Huber, B.\ 2015, arXiv:1505.02782 



\bibitem[\protect\citeauthoryear{Vink \& Laming}{2003}]{vink2003}
Vink, J. \& Laming, J. M. 2003, ApJ, 584, 758

\bibitem[Watson et al.(2013)]{watson2013} 
Watson, W.~A., Iliev, I.~T., D'Aloisio, A., et al.\ 2013, MNRAS, 433, 1230 

\bibitem[Waxman \& Loeb(2000)]{waxman2000} 
Waxman, E., \& Loeb, A.\ 2000, ApJL, 545, L11 

\bibitem[Williamson et al.(2011)]{williamson2011} 
Williamson, R., Benson, B.~A., High, F.~W., et al.\ 2011, ApJ, 738, 139 

\bibitem[White(1994)]{white1994}
White, S. D. M. 1994, arXiv:astro-ph/9410043 

\bibitem[Yamada \& Kitayama(2005)]{yamada2005} 
Yamada, K., \& Kitayama, T.\ 2005, PASJ, 57, 611 


\bibitem[Zandanel \& Ando(2014)]{zandanel2014} 
Zandanel, F., \& Ando, S.\ 2014, MNRAS, 440, 663 




%
\end{thebibliography}
\end{document}